\documentclass[aps,nofootinbib,prd,superscriptaddress,tightenlines,notitlepage,twocolumn,showpacs]{revtex4-1}

\usepackage[colorlinks=true, linkcolor=blue, citecolor=blue]{hyperref}
\usepackage{tabularx}
\usepackage{graphicx}
\usepackage{amssymb,bm}
\usepackage{xcolor}
\usepackage{amsmath}
\usepackage[varg]{txfonts}
\usepackage{enumerate}
\usepackage[normalem]{ulem}
\usepackage[all]{hypcap}

\newcommand{\apjl}{Astrophys. J. Lett.}

\newcommand{\aap}{Astron. \& Astrophys.}

\newcommand{\araa}{Ann. Rev. Astron. Astrophys.} 
\newcommand{\mnras}{Mon. Not. R. Astron. Soc.}
\newcommand{\jcap}{JCAP}

\newcommand{\aapr}{Astron. Astrophys. Rev.}

\newcommand{\PMO}{\affiliation{Purple Mountain Observatory, Chinese Academy of Sciences, Nanjing 210023, China}}
\newcommand{\USTC}{\affiliation{School of Astronomy and Space Sciences, University of Science and Technology of China, Hefei 230026, China}}
\newcommand{\CSIRO}{\affiliation{CSIRO Space and Astronomy, Australia Telescope National Facility, PO Box 76, Epping, NSW 1710, Australia}}

\begin{document}


\title{Bounding the photon mass with gravitationally lensed fast radio bursts}

\author{Chen-Ming Chang}\thanks{changcm@pmo.ac.cn}\PMO\USTC
\author{Jun-Jie Wei}\thanks{jjwei@pmo.ac.cn}\PMO\USTC
\author{Ke-Lai Meng}\thanks{mkl@pmo.ac.cn}\PMO
\author{Song-Bo Zhang}\PMO\CSIRO
\author{Hao-Xuan Gao}\PMO
\author{Jin-Jun Geng}\PMO\USTC
\author{Xue-Feng Wu}\thanks{xfwu@pmo.ac.cn}\PMO\USTC

\date{\today}

\begin{abstract}
    The gravitational time delays of macro-lenses can be used to constrain the rest mass of the photon with high accuracy. 
    Assuming a point-mass $+$ external shear lens model, we prove that an upper limit of the photon mass
    can be derived directly from two observables--the time delay $\Delta t$ and the leading-to-trailing 
    flux ratio $R$ of strongly lensed fast radio bursts (FRBs). Using the observed values of $\Delta t$
    and $R$ of a lensed FRB candidate, i.e., FRB 20190308C, as a reference, we obtain a strict upper limit 
    of the photon mass between $m_\gamma < 5.3 \times {10}^{-42}\,\rm kg$, for a given external shear strength of $\gamma' = 0.01$,
    and $m_{\gamma} < 2.1 \times 10^{-41}-2.4 \times 10^{-42}\,\text{kg}$, within the external shear range of $0<\gamma'<1$.
    This provides the most stringent limit to date on the photon mass through gravitational lensing time delays, 
    improving by 1 to 2 orders of magnitude the previous results obtained from lensed active galactic nuclei.
\end{abstract}

\maketitle


\section{\label{sec:introduction}Introduction}

As one of the fundamental postulates of Maxwell's electromagnetism and Einstein's special relativity, the principle of
invariance of light speed implies that the rest mass of the photon should be exactly zero. Nevertheless, there exist
some theories involving a finite photon rest mass, such as the famous de Broglie-Proca theory \citep{De1922,Proca1936},
the model of the nonvanishing photon mass as an explanation of dark energy \citep{2016PhRvD..93h3012K}, and other new
ideas in the Standard-Model Extension with effectively massive photons \citep{2021EPJC...81....4S}. Despite the great
success of the postulate of the constancy of light speed, those new theories with massive photons are interesting
and worthy to explore, whereas the ultimate word on the photon mass ($m_\gamma$) stems from empirical facts.

Over the last few decades, various kinds of experimental approaches have been performed to push the empirical boundary on the 
masslessness of photons (see \cite{1971RvMP...43..277G,2005RPPh...68...77T,1997ASTPS...4.....Z,Okun2006,Goldhaber2010,
Spavieri2011,2021FrPhy..1644300W} for reviews). These experiments include measurements of the frequency dependence of 
the speed of light ($m_{\gamma}\le3.8\times10^{-51}\,\mathrm{kg}$; \cite{1964Natur.202..377L,1969Natur.222..157W,
1999PhRvL..82.4964S,2016JHEAp..11...20Z,2017RAA....17...13W,2018JCAP...07..045W,2016ApJ...822L..15W,2016PhLB..757..548B,
2017PhLB..768..326B,2017PhRvD..95l3010S,2019ApJ...882L..13X,2020RAA....20..206W,2021PhLB..82036596W,Chang2023,Lin2023,
Wang2023,Ran2024}), tests of Coulomb's inverse square law ($m_{\gamma}\le1.6\times10^{-50}\,\mathrm{kg}$;
\cite{1971PhRvL..26..721W}), measurement of Jupiter's magnetic field ($m_{\gamma}\le8\times10^{-52}\,\mathrm{kg}$;
\cite{1975PhRvL..35.1402D}), analysis of the mechanical stability of magnetized gas in galaxies
($m_{\gamma}\le3\times10^{-63}\,\mathrm{kg}$; \cite{Chibisov1976}), tests of Amp\`ere's law 
($m_{\gamma}\le1.1\times10^{-51}-2.8\times10^{-51}\,\mathrm{kg}$; \cite{1992PhRvL..68.3383C,2024EPJP..139..551S}), magnetohydrodynamics 
of the solar wind ($m_{\gamma}\le1.4\times10^{-49}-3.4\times10^{-51}\,\mathrm{kg}$; 
\cite{1997PPCF...39A..73R,2007PPCF...49..429R,2016APh....82...49R}), Cavendish torsion balance methods 
($m_{\gamma}\le1.2\times10^{-54}\,\mathrm{kg}$; \cite{1998PhRvL..80.1826L,2003PhRvL..90h1801L}), 
estimates of suppermassive black-hole spin ($m_{\gamma}\le7\times10^{-56}\,\mathrm{kg}$; \cite{2012PhRvL.109m1102P}), 
analysis of pulsar spindown ($m_{\gamma}\le6.3\times10^{-53}\,\mathrm{kg}$; \cite{2017ApJ...842...23Y}), 
gravitational deflection of massive photons ($m_{\gamma}\le1.7\times10^{-40}-4.1\times10^{-45}\,\mathrm{kg}$; 
\cite{1973PhRvD...8.2349L,2004PhRvD..69j7501A,2012SCPMA..55..523Q,2014MNRAS.437L..90E,2017ApJ...850..102G}), and so on. 
Among these experiments on photon mass, the resulting constraints obtained from the gravitational deflection of light are not 
the tightest ones; however, in view of model dependence of many experimental methods (see e.g. 
\cite{2005RPPh...68...77T,Goldhaber2010}), tests of the photon mass using different independent methods 
(such as gravitational deflection) are always interesting and important.

The semi-classical gravity predicts that the deflection of massive photons in an external gravitational field 
would be energy-dependent \citep{1973PhRvD...8.2349L,2000PThPh.104..103A,2002CQGra..19.5429A,2004PhRvD..69j7501A}.
Therefore, an upper bound on the photon mass can be obtained by comparing the difference between the measured
deflection angle and the calculated deflection angle for massless photons \citep{1973PhRvD...8.2349L}.
Exploiting the gravitational deflection of radio waves by the Sun, \citet{2004PhRvD..69j7501A} obtained 
an upper limit of $m_\gamma \le {10}^{-43}\,\mathrm{kg}$. Based on the astrometry of several strong gravitational 
lensing systems, \citet{2012SCPMA..55..523Q} investigated the photon mass limit at a cosmological scale, yielding 
$m_\gamma \le 8.7\times {10}^{-42}\,\mathrm{kg}$. With the precise astrometry of the gravitationally lensed quasar 
MG J2016$+$112, \citet{2014MNRAS.437L..90E} further improved the limit to be $m_\gamma \leq 4.1 \times {10}^{-45}\,\rm kg$.
However, these astrometric limits do not use lens models and simply assume that the angular separation of  
lensed images is equivalent to the deflection angle of light. \citet{2017ApJ...850..102G} argued that this is 
a strong assumption. Modeling lens galaxies with central suppermassive black holes by a singular isothermal model,
\citet{2017ApJ...850..102G} used the time delays between compact images from three lensed active galactic nuclei 
(AGNs) to derive a photon mass limit of $m_\gamma \leq 1.7 \times {10}^{-40}\,\rm kg$.

Fast radio bursts (FRBs) are bright millisecond-long radio flashes originating at cosmological distances
\citep{2007Sci...318..777L,2013Sci...341...53T,2019ARA&A..57..417C,2019A&ARv..27....4P,2022A&ARv..30....2P,2023RvMP...95c5005Z}.
Their cosmological origin, energetic nature, and high all-sky rate make them ideal for probing cosmology
and fundamental physics (e.g., \cite{Deng2014}). Indeed, FRBs have been widely used to constrain 
the photon mass through measurements of the frequency dependence of the speed of light \citep{2016ApJ...822L..15W,2016PhLB..757..548B,2017PhLB..768..326B,2017PhRvD..95l3010S,2019ApJ...882L..13X,2020RAA....20..206W,2021PhLB..82036596W,Chang2023,Lin2023,
Wang2023,Ran2024}. With tens of thousands of signals that will be guaranteed in the future, FRBs 
have gained attention as potential targets for lensing studies \citep{2016PhRvL.117i1301M,2018A&A...614A..50W,
2020ApJ...896L..11L,2020ApJ...900..122S,2022PhRvD.105j3528K,2022PhRvD.106d3017L,2022ApJ...928..124Z,2023MNRAS.521.4024C}. 
Very recently, \citet{2024arXiv240619654C} employed the autocorrelation algorithm to search for potential 
lensed FRBs in the first Canadian Hydrogen Intensity Mapping Experiment (CHIME) FRB catalogue, and identified 
FRB 20190308C as a lensed candidate with a significance of $3.4\sigma$. The information about the time delay
and flux ratio between the two substructures of FRB 20190308C can be easily extracted. Inspired by 
\citet{2017ApJ...850..102G}, a natural question arises: is it possible to improve the photon mass lensing limits
by using the gravitational time delays of lensed FRBs?

In this work, we propose a new method to place an upper limit on the photon mass by applying the time delay 
information from lensed FRBs. The rest of this paper is arranged as follows. In Section~\ref{sec:method}, 
we discuss the photon-mass dependence of the time delay in the Chang-Refsdal lens model.
The constraints on the photon mass from a lensed FRB candidate are presented in Section~\ref{sec:result}. 
Finally, a brief summary and discussion are provided in Section~\ref{sec:conclusion}.

\section{\label{sec:method}Photon-Mass Dependence of the Time Delay in the Chang-Refsdal Lens Model}

The Chang-Refsdal lens model describes the lensing effect of a star, which can be considered as a point-mass lens under 
the gravitational perturbation of a background galaxy. The lens equation of the Chang-Refsdal lens model is given by 
\citep{1979Natur.282..561C,1984A&A...132..168C,2006MNRAS.369..317A,2021ApJ...912..134C,2022MNRAS.516..453G}
\begin{equation}\label{eq:lens-equation}
    \begin{aligned}
        \beta_1 -\theta_1 &=   \gamma\theta_1-\theta_E^2\frac{\theta_1}{{\left|\theta\right|}^2},\\
        \beta_2 - \theta_2 &=  -\gamma\theta_2-\theta_E^2\frac{\theta_2}{{\left|\theta\right|}^2},
    \end{aligned}
\end{equation}
where $\beta$ and $\theta$ represent the positions in the source and deflector planes, respectively, 
$\theta_E$ is the Einstein angle, and $\gamma$ is the external shear strength. Note that $\alpha=\beta -\theta$
stands for the deflection angle of light.

Assuming a weak gravitational field, the basic formulas for the time delay and position of lensed images of a massive photon
source were derived by \citet{1973PhRvD...8.2349L} and \citet{2017ApJ...850..102G}. These studies show that the deflection angle
for massive photons is similar as that of massless photons, except for the $(1+\frac{1}{2}\mu^2)$ multiplicative factor.
Here $\mu^2 = \frac{{m_\gamma}^2c^2}{P_0^2}$, where $m_\gamma$ is the rest mass of the photon and $P_0$ is the time component 
of the four-momentum. Therefore, for the scenario of massive photons, the lens equation (Equation~\ref{eq:lens-equation}) 
can be simply rewritten by replacing $\gamma$ and $\theta_E^2$ with $\gamma'=(1+\frac{1}{2}\mu^2)\gamma$ and
${\theta_E'}^2=(1+\frac{1}{2}\mu^2)\theta_E^2$. Scaling the angular coordinates with $\theta_E'$: $y'=\beta/\theta_E'$ and
$x'=\theta/\theta_E'$, the dimensionless lens equation reads as
\begin{equation}\label{lensing equation}
    \begin{aligned}
        y_1' &= (1+\gamma')x_1'-\frac{x_1'}{{x_1'}^2+{x_2'}^2},\\
        y_2' &= (1-\gamma')x_2'-\frac{x_2'}{{x_1'}^2+{x_2'}^2}.
    \end{aligned}
\end{equation} 
This lens equation can have up to four solutions, resulting in multiple imaging scenarios. It is difficult to obtain 
the general analytical solutions to the lens equation, so do the expressions for the time delay $\Delta t$ and 
flux ratio $R$ between lensed images. Fortunately, the ``permitted region'' in the $R$--$\Delta t$ space can be determined 
with the three boundary conditions that the source is on the symmetry axis of the lensing system 
(i.e., $y_1' = 0$ or $y_2' = 0$) or at the tips of the inner caustics, thereby providing bounds on the photon mass.

\citet{2021ApJ...912..134C} focused on two-image configurations with $\gamma \ll 1$. For the case of $\gamma \ll 1$, the size of caustic is much less than $\theta_E'$ and the cross section of four-image configurations can be ignored. Their analysis can be extended to the case of $\gamma < 1$ if we only consider two-image configurations.

For $y_1' = 0$, the lower boundary of the permitted region in the $R$--$\Delta t$ space can be determined. 
The solutions for Equation (\ref{lensing equation}) are 
\begin{equation}\label{sol1_y1=0}
    \begin{aligned}
        x_1' &= 0,\\
        x_2' &= \frac{y_2' \pm \sqrt{{y_2'}^2+4\left(1-\gamma'\right)}}{2\left(1-\gamma'\right)};
    \end{aligned}
\end{equation} 
and
\begin{equation}\label{sol2_y1=0}
    \begin{aligned}
        x_1' &= \pm \sqrt{\frac{1}{1+\gamma'}-\frac{{y_2'}^2}{4{\gamma'}^2}},\\
        x_2' &= -\frac{y_2'}{2\gamma'}.
    \end{aligned}
\end{equation} 
Using Equation~(\ref{sol1_y1=0}) and the magnification of each image (see \citet{2021ApJ...912..134C} 
for the detailed derivation), we obtain the time delay  
\begin{equation}\label{t_y1=0}
    \Delta t = \frac{4GM}{c^3}\left(1+z_l\right)\left[\frac{y_2's_2'}{2\left(1-\gamma'\right)}+\ln\left(\frac{s_2'+y_2'}{s_2'-y_2'}\right)\right],
\end{equation} 
and the flux ratio
\begin{equation}\label{R_y1=0}
    R = \frac{\left({y_2'}^2+2+y_2's_2'\right)\left(y_2's_2'-4\gamma'\right)+8{\gamma'}^2+2\gamma'{y_2'}^2}{\left({y_2'}^2+2-y_2's_2'\right)\left(y_2's_2'+4\gamma'\right)-8{\gamma'}^2-2\gamma'{y_2'}^2},
\end{equation} 
where $M$ and $z_l$ are the point mass and redshift of the lens, respectively, and $s_2' = \sqrt{{y_2'}^2+4(1-\gamma')}$.

For $y_2' = 0$, the solutions for Equation (\ref{lensing equation}) are
\begin{equation}\label{sol1_y2=0}
    \begin{aligned}
        x_1' &= \frac{y_1' \pm \sqrt{{y_1'}^2+4\left(1+\gamma'\right)}}{2\left(1+\gamma'\right)},\\
        x_2' &= 0;
    \end{aligned}
\end{equation} 
and
\begin{equation}\label{sol2_y2=0}
    \begin{aligned}
        x_1' &= \frac{y_1'}{2\gamma'},\\
        x_2' &= \pm \sqrt{\frac{1}{1-\gamma'}-\frac{{y_1'}^2}{4{\gamma'}^2}}.
    \end{aligned}
\end{equation} 
When $y_2' = 0$, the corresponding formulas for the time delay and flux ratio can be treated as the upper boundary of the permitted region, i.e.,
\begin{equation}\label{t_y2=0}
    \Delta t = \frac{4GM}{c^3}\left(1+z_l\right)\left[\frac{y_1's_1'}{2\left(1+\gamma'\right)}+\ln\left(\frac{s_1'+y_1'}{s_1'-y_1'}\right)\right],
\end{equation} 
and 
\begin{equation}\label{R_y2=0}
    R = \frac{\left({y_1'}^2+2+y_1's_1'\right)\left(y_1's_1'+4\gamma'\right)+8{\gamma'}^2-2\gamma'{y_1'}^2}{\left({y_1'}^2+2-y_1's_1'\right)\left(y_1's_1'-4\gamma'\right)-8{\gamma'}^2+2\gamma'{y_1'}^2},
\end{equation} 
where $s_1' = \sqrt{{y_1'}^2+4(1+\gamma')}$.

When the source is located at the tips of the inner caustics, i.e., $y_1' = 0$ and $y_2'=\pm2\gamma'/\sqrt{1+\gamma'}$
(or $y_1'=\pm2\gamma'/\sqrt{1-\gamma'}$ and $y_2' = 0$), the left boundary of the permitted region in the $R$--$\Delta t$ 
space can be determined. The lower limit on $\Delta t$ can be written as 
\begin{equation}\label{mint}
    \Delta t_{\rm min} = \frac{4GM}{c^3}\left(1+z_l\right)\left[\frac{2\gamma'}{1-{\gamma'}^2}+\ln\left(\frac{1+\gamma'}{1-\gamma'}\right)\right].
\end{equation}
The corresponding leading-to-trailing flux ratio $R$ is $0$ or $+\infty$.

These boundary conditions can still provide some information about the mass of the lens and 
the photon-mass-dependent time delay between lensed images in the absence of the general analytical solutions. 
Therefore, they can be used for further study on photon mass limits.

\section{\label{sec:result}Photon Mass Limit from A Lensed FRB Candidate}

\begin{figure}
    \includegraphics[width=\columnwidth]{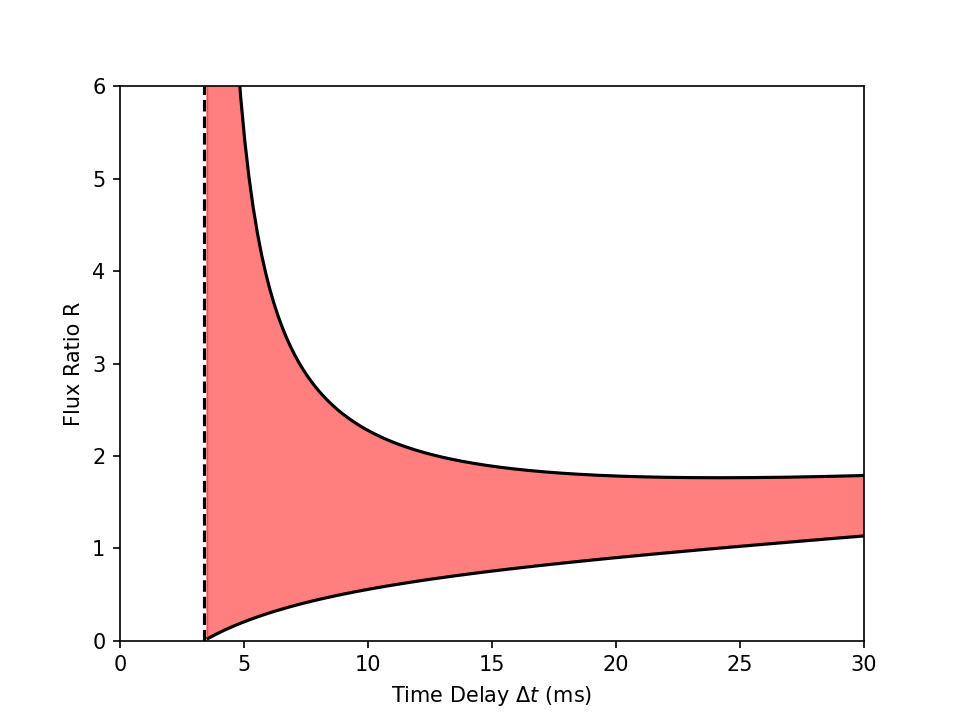}
    \caption{Distributions of the leading-to-trailing flux ratio $R$ and time delay $\Delta t$ 
    for a point-mass $+$ external shear lens model. The red shaded area represents the permitted 
    region of all possible $R$--$\Delta t$ pairs with the lens mass $M(1+z_l) = 4277\, {\rm M}_\odot$ 
    and the equivalent external shear strength $\gamma' = 0.01$. The vertical dashed line and 
    two black solid curves correspond to the left, upper, and lower boundaries of all possible 
    $R$--$\Delta t$ pairs.}
    \label{fig:permissible_region}
\end{figure}

The permitted region of all possible $R$--$\Delta t$ pairs for a point-mass $+$ external shear lens model
with the lens mass $M(1+z_l)=4277\,{\rm M}_\odot$ and the equivalent external shear strength 
$\gamma' = 0.01$\footnote{Note that for the case of $\gamma' = 0.01$, $4277\,{\rm M}_\odot$
is the lower lens mass bound for a lensed FRB candidate (FRB 20190308C), as discussed in \citet{2024arXiv240619654C}. 
Further details are provided below.} is shown in 
Figure~\ref{fig:permissible_region}. One can see from this plot that all possible $R$--$\Delta t$ pairs between two
lensed images are actually bracketed by three boundary lines. The lower boundary of the permitted region (solid curve 
on bottom) corresponds to the $R$--$\Delta t$ relation along the $y_2'$-axis (i.e., $y_1' = 0$), which is determined by 
Equations~(\ref{t_y1=0}) and (\ref{R_y1=0}). The upper boundary (solid curve on top) corresponds to the $R$--$\Delta t$ 
relation along the $y_1'$-axis (i.e., $y_2' = 0$), which is determined by Equations~(\ref{t_y2=0}) and (\ref{R_y2=0}).
This $y_2' = 0$ curve reaches its minimum at $y_1'=\sqrt{\frac{2\gamma'(1+2\gamma')}{1-\gamma'}}$ with
\begin{equation}
\begin{split}
    &R_{\rm min} = \\ &\frac{\left(4{\gamma'}^2+2+\Delta'\right)\left(\Delta'+4\gamma'-4{\gamma'}^2\right)+16{\gamma'}^4-20{\gamma'}^3+4{\gamma'}^2}{\left(4{\gamma'}^2+2-\Delta'\right)\left(\Delta'-4\gamma'+4{\gamma'}^2\right)-16{\gamma'}^4+20{\gamma'}^3-4{\gamma'}^2},
\end{split}
\end{equation}
where $\Delta' = \sqrt{8{\gamma'}^3+20{\gamma'}^2+8\gamma'}$. It is obvious that $R_{\rm min}$ is larger than $1$
when $\gamma' > 0$. The left boundary (vertical dashed line) corresponds to the lower limit of $\Delta t$, which is 
determined by Equation~(\ref{mint}). As shown in Figure~\ref{fig:permissible_region}, with the fixed $R$, 
the observed time delay $\Delta t_{\rm obs}$ between the lensed images should always be larger than the lower limit 
$\Delta t_{\rm min}$ (vertical dashed line). With Equation~(\ref{mint}), it is thus easy to obtain 
\begin{equation}
    \frac{4GM}{c^3}\left(1+z_l\right)\cdot4\left(1+\frac{1}{2}\mu^2\right)\gamma < \Delta t_{\rm min} < \Delta t_{\rm obs}.
\end{equation} 
So the photon mass can be constrained as
\begin{equation}\label{bound}
    m_\gamma < \frac{P_0}{c}\sqrt{\frac{c^3\Delta t_{\rm obs}}{8GM\left(1+z_l\right)\gamma}-2}.
\end{equation}

\begin{figure}
    \includegraphics[width=\columnwidth]{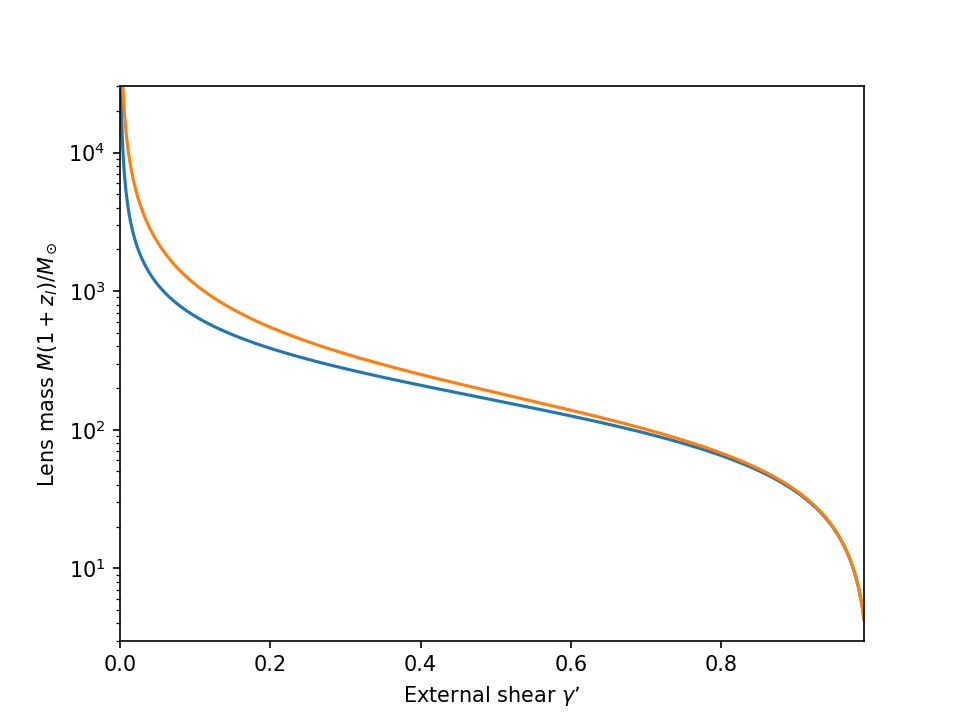}
    \caption{$M(1+z_l)$--$\gamma'$ relations that correspond to the situation of $R_{\rm obs} = 0.5$
and $\Delta t_{\rm obs} = 8.85\,\rm ms$. The orange curve represents the upper limit of $M(1+z_l)$ obtained with Equation (\ref{mint}). The blue curve represents the lower limit of $M(1+z_l)$ obtained with Equations (\ref{t_y1=0}) and (\ref{R_y1=0}).}
    \label{fig:mass_range}
\end{figure}

Very recently, \citet{2024arXiv240619654C} searched for potential lensed FRBs within the first CHIME/FRB catalogue
using the autocorrelation algorithm and verification through signal simulations. Only FRB 20190308C was identified 
as a plausible candidate for gravitational lensing. The observed time delay and flux ratio between the two substructures 
of FRB 20190308C are $\Delta t_{\rm obs} = 8.85\,\rm ms$ and $R_{\rm obs} = 0.5$. As an example, we now use 
the time delay information from FRB 20190308C to demonstrate how to obtain the constraints on the lens mass $M(1+z_l)$,
thereby placing constraints on the photon mass $m_\gamma$. For the doubly lensed FRB 20190308C with $R_{\rm obs} = 0.5$
and $\Delta t_{\rm obs} = 8.85\,\rm ms$,
the upper boundary does not provide any useful information because $0.5$ will always be smaller than $R_{\rm min}$
for any $\gamma'$. Therefore, only two $M(1+z_l)$--$\gamma'$ relations derived from the lower and left boundaries
are seen in Figure~\ref{fig:mass_range}. The $M(1+z_l)$--$\gamma'$ relation corresponding to the lower boundary 
(blue line) is derived from Equations~(\ref{t_y1=0}) and (\ref{R_y1=0}). The $M(1+z_l)$--$\gamma'$ relation 
corresponding to the left boundary (orange line) is derived from Equation~(\ref{mint}). For a moderate shear of
$\gamma' = 0.01$, the lower limit on the lens mass is about $M(1+z_l)=4277\,{\rm M}_\odot$ (see \cite{2024arXiv240619654C} for more details).
Since the photon mass term $\mu \ll 1$, it is reasonable to assume that $\gamma \simeq \gamma' = 0.01$. 
With the observed time delay $\Delta t_{\rm obs} = 8.85\,\rm ms$, the lowest observed frequency 
$\nu = \frac{P_0c}{h} = 400\,\rm MHz$, and the lower lens mass limit $M(1+z_l)=4277\,{\rm M}_\odot$ 
corresponding to $\gamma' = 0.01$, a stringent upper limit on the photon mass from Equation~(\ref{bound}) is
\begin{equation}
    m_\gamma < 5.3 \times {10}^{-42}\,\rm kg
\end{equation}
for FRB 20190308C.

\begin{figure}
    \includegraphics[width=\columnwidth]{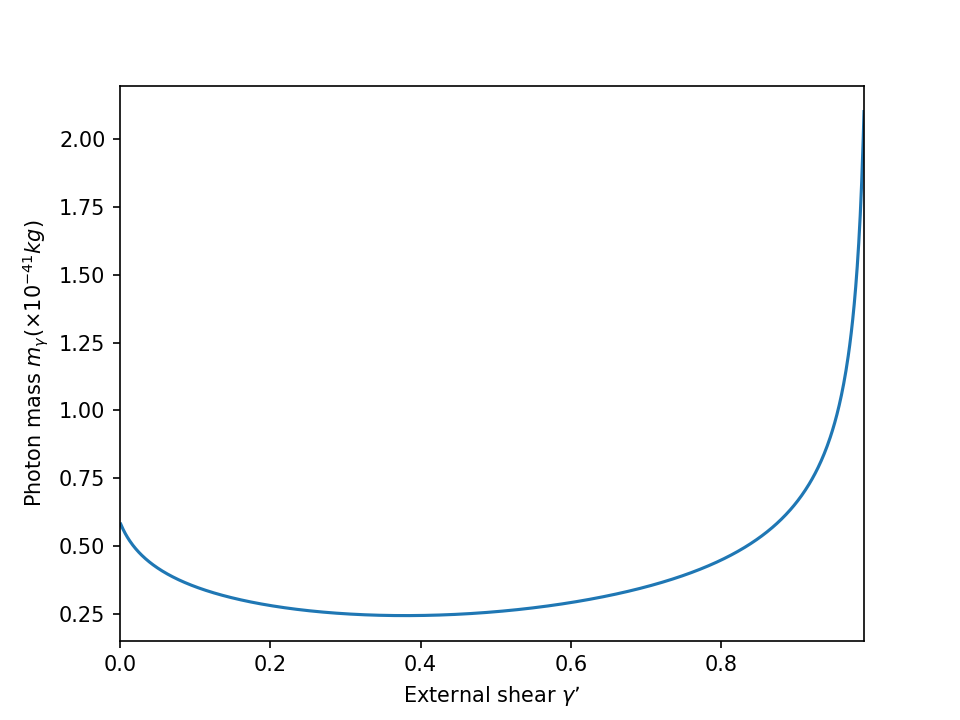}
    \caption{Sensitivity of the upper photon mass limit $m_\gamma$ to the equivalent external shear strength $\gamma'$.}
    \label{fig:photon_mass}
\end{figure}

In our above analysis, the external shear strength is set to be $\gamma'=0.01$. To explore 
the effect of different $\gamma'$ values, we estimate the sensitivity as we vary $\gamma'$.
For $\gamma' < 1$, the solutions (i.e., Equation~\ref{sol1_y1=0}) for the dimensionless lens 
equation always exist when $y_1' = 0$. Therefore, $\left|y_2'\right|$ needs to be greater than
$\frac{2\gamma'}{\sqrt{1+\gamma'}}$ to satisfy the two-image condition and ensure that the
boundary conditions discussed in Section \ref{sec:method} remain applicable.
As shown in Figure~\ref{fig:mass_range}, with a fixed $\gamma'$, a lower limit on the lens mass limit $M(1+z_l)$
can be obtained, leading to the establishment of an upper limit on the photon mass $m_\gamma$.
Figure~\ref{fig:photon_mass} shows that as $\gamma'$ increases, $m_\gamma$ decreases first and then increases.
The upper limit of the photon mass has a minimum value of $m_{\gamma} = 2.4 \times 10^{-42}\,\text{kg}$, 
corresponding to $\gamma' = 0.38$ and a minimum lens mass of $M(1+z_l) = 221.7 \,{\rm M}_\odot$. 
The maximum value of the upper photon mass limit is difficult to determine because, as $\gamma'$ approaches 1, 
the wave properties of light become significant, making the geometric approximation of the lensing equation invalid. 
Therefore, we calculate the upper limit of the photon mass to be $m_{\gamma} < 2.1 \times 10^{-41}\,\text{kg}$ 
when $\gamma' = 0.99$, corresponding to a minimum lens mass of $M(1+z_l) = 4.3 \,{\rm M}_\odot$.
That is, within the range of $0<\gamma'<1$, the photon mass can be constrained to be 
\begin{equation}
    m_{\gamma} < 2.1 \times 10^{-41}-2.4 \times 10^{-42}\,\text{kg}
\end{equation}
for FRB 20190308C, which is almost 10-100 times tighter than the constraints from the time delays of lensed AGNs
\citep{2017ApJ...850..102G}.

For $\gamma' > 1$, there are two caustics, each with one cusp on the $y_2'$ axis and two cusps off-axis. The boundary conditions for the two-image scenario are not clear. When $y_1' = 0$, the two-image scenario has two possibilities: either $\left|y_2'\right| \geq \frac{2\gamma'}{\sqrt{1+\gamma'}}$ or $\left|y_2'\right| \leq 2\sqrt{\gamma'-1}$. For $y_1' = 0$ and $\left|y_2'\right| \geq \frac{2\gamma'}{\sqrt{1+\gamma'}}$, the time delay and flux ratio become: 
\begin{equation}\label{deltat,y1=0}
    \Delta t = \frac{4GM}{c^3}\left(1+z_l\right)\left[\frac{y_2's_2'}{2\left(\gamma'-1\right)}-\ln\left(\frac{y_2'+s_2'}{y_2'-s_2'}\right)\right],
\end{equation} 
and 
\begin{equation}\label{R,y1=0}
    R = \frac{-\left({y_2'}^2+2-y_2's_2'\right)\left(y_2's_2'+4\gamma'\right)+8{\gamma'}^2+2\gamma'{y_2'}^2}{\left({y_2'}^2+2+y_2's_2'\right)\left(y_2's_2'-4\gamma'\right)+8{\gamma'}^2+2\gamma'{y_2'}^2}.
\end{equation} 
For $y_1' = 0$ and $\left|y_2'\right| \leq 2\sqrt{\gamma'-1}$, $\Delta t = 0$ and $R = 1$. Numerical calculations suggest that it may still be possible to impose certain constraints on the combined lens mass $M(1+z_l)$ using Equations~(\ref{deltat,y1=0}) and (\ref{R,y1=0}) by adding some restrictions, such as limiting the leading-to-trailing flux ratio $R < 1$. However, the information obtained from the dynamic spectrum may not be sufficient to constrain the photon mass $m_\gamma$ due to the lack of an analytical relationship between $M(1+z_l)$ and $m_\gamma$.

\section{\label{sec:conclusion}Summary and Discussion}

It has been suggested that the strong lensing effect of a point mass $+$ external shear lens model on a single-peak FRB
can produce double peaks (i.e., lensed images). Based on this lens model, here we proposed a method of using the two observables 
of the time delay $\Delta t$ and the leading-to-trailing flux ratio $R$ from lensed FRBs to set a stringent upper limit
on the photon mass. In particular, we showed the process of constraining photon mass using the observed values of 
$\Delta t$ and $R$ from a lensed FRB candidate, i.e., FRB 20190308C, as a reference. 

For a point mass $+$ external shear lens model, there is no one-to-one correspondence between the upper photon mass
limit $m_\gamma$ and $\Delta t$ and $R$ due to the extra freedom of the external shear. Nevertheless, we showed that
an upper limit on $m_\gamma$ can still be derived from $\Delta t$ and $R$ for a given external shear strength of $\gamma'$
(Section~\ref{sec:result}). For FRB 20190308C with $\Delta t = 8.85\,\rm ms$ and $R = 0.5$, we obtained a strict constraint 
on the photon mass $m_\gamma < 5.3 \times {10}^{-42}\,\rm kg$ for a fixed external shear strength of $\gamma' = 0.01$.
We also inspected the influences of different $\gamma'$ values, finding that this effect has a modest impact on 
the photon mass limits. That is, within the range of $0<\gamma'<1$, one can derive $m_{\gamma} < 2.1 \times 10^{-41}-2.4 \times 10^{-42}\,\text{kg}$.

Previously, by analyzing the gravitational time delays from lensed AGNs, \citet{2017ApJ...850..102G}
set a severe limit on the photon mass of $m_\gamma \leq 1.7 \times {10}^{-40}\,\rm kg$.
In the present Letter, using the sharp features of the lensed FRB signals, we have obtained 
the most stringent limit to date on the photon mass through gravitational lensing time delays, 
namely $\sim 2.1 \times 10^{-41}-2.4 \times 10^{-42}\,\text{kg}$, which represents an improvement of 
1 to 2 orders of magnitude over the results previously obtained from lensed AGNs.

So far, only a lensed FRB candidate with a significance of $3.4\sigma$ has been identified \cite{2024arXiv240619654C}.
Nevertheless, given the high all-sky event rate \cite{2013Sci...341...53T,2023MNRAS.521.4024C} and sustained efforts in FRB searches, more FRB
signals lensed by point-mass lenses with higher significance are expected to be identified in the near future.
Due to the short-lived nature and unpredictability of FRBs, it may be hard to perform a full lens modeling
with the observed data. The method presented in this work offers an alternative, straightforward way of
constraining the photon mass from easily obtained observables of $\Delta t$ and $R$.

\begin{acknowledgments}
We are grateful to the anonymous referee for the helpful comments.
This work is supported by the National SKA Program of China (2022SKA0130100), 
the National Natural Science Foundation of China (grant Nos. 12422307, 12373053, 12321003, and
12041306), the Key Research Program of Frontier Sciences (grant No. ZDBS-LY-7014)
of Chinese Academy of Sciences, International Partnership Program of Chinese Academy of Sciences
for Grand Challenges (114332KYSB20210018), the CAS Project for Young Scientists in Basic Research
(grant No. YSBR-063), and the Natural Science Foundation of Jiangsu Province (grant No. BK20221562).
\end{acknowledgments}

\bibliographystyle{apsrev4-1}
\bibliography{apssamp}

\end{document}